\documentclass{jkas}

\def\year{2017} 
\def\month{November} 
\def\volume{00} 
\def\issue{0} 
\def\beginpage{1} 
\def\endpage{12} 
\def\received{May 23, 2017} 
\def\accepted{November 16, 2017} 
\date{Received \received; accepted \accepted}

\usepackage{flushend} 
\usepackage{microtype}
\usepackage{xcolor}

\usepackage[pdftex, breaklinks, colorlinks, citecolor=blue, linkcolor=blue, menucolor=blue, urlcolor=blue]{hyperref}

\title{
The Millimeter-Radio Emission of BL Lacertae\\ During Two $\mathbf\gamma$-ray Outbursts
}

\author[1]{Dae-Won~Kim}
\author[1]{Sascha~Trippe}
\author[2,3]{Sang-Sung~Lee}
\author[1]{Jong-Ho~Park}
\author[4]{Jae-Young~Kim}
\author[1]{Juan-Carlos~Algaba}
\author[2]{Jeffrey~A.~Hodgson}
\author[5,6]{Motoki~Kino}
\author[2]{Guang-Yao~Zhao}
\author[2]{Kiyoaki~Wajima}
\author[2,3]{Sincheol~Kang}
\author[1]{Junghwan~Oh}
\author[1]{Taeseok~Lee}
\author[2,3]{Do-Young~Byun}
\author[2,3]{Soon-Wook~Kim}
\author[5]{Jeong-Sook~Kim}

\affil[1]{Department of Physics and Astronomy, Seoul National University, Gwanak-gu, Seoul 08826, Korea; \email{dwkim@astro.snu.ac.kr, trippe@astro.snu.ac.kr}}
\affil[2]{Korea Astronomy and Space Science Institute, 776 Daedeok-daero, Yuseong-gu, Daejeon 34055, Korea}
\affil[3]{Korea University of Science and Technology, 217 Gajeong-ro, Yuseong-gu, Daejeon 34113, Korea}
\affil[4]{Max-Planck-Institut f\"{u}r Radioastronomie, Auf dem H\"{u}gel 69, 53121 Bonn, Germany}
\affil[5]{National Astronomical Observatory of Japan, 2-21-1 Osawa, Mitaka, Tokyo, 181-8588, Japan}
\affil[6]{Kogakuin University, Academic Support Center,
2665-1 Nakano, Hachioji, Tokyo 192-0015, Japan}

\def\corrauthor{
S. Trippe
}

\def\runningauthor{
Kim et al.
}

\def\runningtitle{
The Millimeter-Radio Emission of BL Lacertae During Two $\gamma$-ray Outbursts
}

\def\keywords{
galaxies: active --- BL Lacertae objects: individual (BL Lacertae) --- galaxies: jets --- techniques: interferometric --- radio continuum: galaxies
}

\def\abstracttext{
  We present a study of the inexplicit connection between radio jet activity and $\gamma$-ray emission of BL Lacertae (BL Lac; 2200+420). We analyze the long-term millimeter activity of BL Lac via interferometric observations with the Korean VLBI Network (KVN) obtained at 22, 43, 86, and 129 GHz simultaneously over three years (from January 2013 to March 2016); during this time, two $\gamma$-ray outbursts (in November 2013 and March 2015) can be seen in $\gamma$-ray light curves obtained from \emph{Fermi} observations. The KVN radio core is optically thick at least up to 86 GHz; there is indication that it might be optically thin at higher frequencies. To first order, the radio light curves decay exponentially over the time span covered by our observations, with decay timescales of 411$\pm$85 days, 352$\pm$79 days, 310$\pm$57 days, and 283$\pm$55 days at 22, 43, 86, and 129 GHz, respectively. Assuming synchrotron cooling, a cooling time of around one year is consistent with magnetic field strengths $B\sim2\,\mu$T and electron Lorentz factors $\gamma\sim10\,000$. Taking into account that our formal measurement errors include intrinsic variability and thus over-estimate the statistical uncertainties, we find that the decay timescale $\tau$ scales with frequency $\nu$ like $\tau\propto\nu^{-0.2}$. This relation is much shallower than the one expected from opacity effects (core shift), but in agreement with the (sub-)mm radio core being a standing recollimation shock. We do not find convincing radio flux counterparts to the $\gamma$-ray outbursts. The spectral evolution is consistent with the `generalized shock model' of \citet{valtaoja1992}. A temporary increase in the core opacity and the emergence of a knot around the time of the second $\gamma$-ray event indicate that this $\gamma$-ray outburst might be an `orphan' flare powered by the `ring of fire' mechanism.
}

\begin{document}
\jkashead 


\section{Introduction\label{sec:1}}

Certain active galactic nuclei (AGN) show prominent jets, powerful plasma outflows from the immediate environment of the central black hole. Jets are luminous emitters of synchrotron radiation which originates from the interplay of relativistic electrons and magnetic fields with complex structure \citep[see, e.g.,][]{hughes1991, bottcher2012}. Propagating disturbances, especially shocks, are supposed to play an important role in the observed variability of jets \citep{hughes1985, marscher2008}. Blazars, a subclass of AGN marked by relativistic outflow speeds, high luminosities, and strong variability, are obvious candidates to explore the nature of AGN jets. Blazar jets are oriented close to the line of sight, thus causing substantial Doppler boosting. Typically, blazar jets radiate non-thermal emission over the wide spectral range from radio to $\gamma$-rays. A typical spectral energy distribution (SED) consists of two `humps': one which is located at relatively low frequencies from radio to UV which is attributed to synchrotron radiation; one at higher frequencies all the way up to $\gamma$-rays which is attributed to inverse Compton (IC) radiation (e.g., \citealt{ramakrishnan2015}). To date, key physical processes in blazar jets, such as the acceleration of relativistic electrons, the structure of the inner jet region, the magnetic field structure, the origin of the seed photons for the IC process, and multi-waveband correlations are still unclear. 

Broadly, blazars can be sub-divided into BL~Lacertae (BL~Lac) objects and flat-spectrum radio quasars (FSRQs), using the strength of broad emission lines, the synchrotron peak frequencies of their SEDs, and the shape of the spectral continuum as distinctive criteria. Even though, all blazar sub-classes can be understood as members of a continuous blazar sequence (\citealt{ghisellini1998}, \citealt{fossati1998}; but see also \citealt{padovani2007}, \citealt{giommi2012}). In this picture, the two peak frequencies of the synchrotron and IC components are connected to each other and show an anti-correlation with bolometric luminosity (the power of the jet) (e.g., \citealt{fossati1998}).

The connection between radio and $\gamma$-ray emission in blazars is a matter of ongoing debate \citep{jorstad2001a, marscher2010, marscher2011, chatterjee2012, raiteri2013, marscher2014, ramakrishnan2016, wehrle2016}. On the one hand, it has been reported that $\gamma$-ray outbursts occur during the rising or peaking state of mm-radio flares (the ``radio counterparts'') \citep{valtaoja1995, tavares2011}. The ejection of new jet components and $\gamma$-ray outbursts are correlated and tend to cause an increase in core opacity \citep{jorstad2001b, jorstad2013}. In addition, \citet{jorstad2001a} found a positive correlation between radio core flux and $\gamma$-ray flux in 42 $\gamma$-ray bright blazars; this suggests that the $\gamma$-ray emission site is located near the core. \citet{tavares2011} divided the vicinity of the radio core into the three regions --  upstream, passing, and downstream of the core -- to locate the $\gamma$-ray production site, and suggests possible scenarios for each case. On the other hand, several studies do not find significant correlations between radio and $\gamma$-ray emission from BL~Lac objects \citep[e.g.,][]{tavares2011} and blazars \citep[e.g.,][]{moerbeck2014}. Hence, the connection between radio and $\gamma$-ray emission from blazar jets is rather obscure. The picture is complicated further by `orphan' $\gamma$-ray flares that have no counterparts at lower frequencies \citep{krawczynski2004, bottcher2005, macdonald2015, banasinski2016}. Orphan $\gamma$-ray flares are thought to occur occasionally in BL Lac objects and a few FSRQs \citep{banasinski2016}. A number of studies suggests possible origins of orphan flares: (i) a sudden enhancement of an external radiation field (e.g., from the accretion disk) \citep{krawczynski2004}; (ii) a slow jet sheath (the `ring of fire') in front of the radio core provides localized seed photons to a relativistic plasma cloud during its passage through the ring \citep{macdonald2015}; (iii) an interaction of relativistic jet particles with a dense radiation field or stellar wind from a luminous star located around the central engine \citep{banasinski2016}.

BL Lacertae (redshift $z=0.069$, image scale 1.29 pc/mas), the prototypical BL~Lac object, is characterized by strong flux variability from radio to $\gamma$-rays, high radio brightness ($>$1~Jy), superluminal motion of jet components, and linearly polarized emission. \citet{villata2004} reported evidence for a (quasi-)periodic occurrence of major radio flares on a timescale of about eight years. BL Lac is known as a $\gamma$-ray bright AGN, beginning with the first detection of $\gamma$-ray emission by \citet{catanese1997} in early 1995. Subsequent detections or reports occurred in 1997, 2005, 2011, 2012, and 2013 by \citet{bloom1997}, \citet{albert2007}, \citet{arlen2013}, \citet{wehrle2016}, and \citet{gomez2016}, respectively. BL Lac was highly active at $\gamma$-ray and radio in 2012, showing a very strong radio outburst that peaked at the end of 2012 \citep[see also][for recent discussions]{gomez2016, wehrle2016}. \citet{gomez2016} detected two polarized stationary features in the jet at 0.1 mas and 0.25 mas from the core, which they interpreted as recollimation shocks. \citet{arlen2013} found a $\gamma$-ray flare in 2011 with a radio counterpart following four months later. Subsequently, \citet{wehrle2016} confirmed that a $\gamma$-ray flare in 2012 is associated with the historic 2012 radio outburst, showing a time-lag of 3--5 months (again with the $\gamma$-ray emission leading the radio emission).

The Korean VLBI Network (KVN) is a dedicated mm-radio VLBI array with the unique capability of observing at four frequencies simultaneously \citep{lee2014}. The KVN covers 22, 43, 86, and 129 GHz in frequency, and consists of three antennas with a maximum baseline length of 476~km. Systematic monitoring of blazar jets  with multi-frequency simultaneous VLBI observations beyond 43 GHz is rare. Hence, the KVN will provide important information on variability, spectral properties, and evolution of blazar jets. We aim at exploring the connection between $\gamma$-ray and radio emission by monitoring the activity of the radio jet of BL~Lac. In this study, we analyze the flux from and structure of BL~Lac by employing dedicated KVN observations spanning three years in time, from January 2013 to March 2016; during this time, two $\gamma$-ray outbursts occured.

In the following, we describe our data and observations in Section~\ref{sec:2}. In Section~\ref{sec:3}, we present the observational results, and we discuss the results in Section~\ref{sec:4}. Lastly, we summarize this paper in Section~\ref{sec:5}.

\begin{table}[t!]
\caption{Summary of observations\label{tab:obslog}}
\centering 
\begin{tabular}{c c c c}
\toprule			
Date & Session & Frequency$^{\rm a}$ & $t_{obs}^{\rm b}$ \\
     &         & (GHz)     & (min)        \\
\midrule
2013-01-16  &  iMOGABA2  &  22/43/86  &  25     \\
2013-02-27  &  iMOGABA3  &  22/43/86/129  &  25    \\
2013-03-28  &  iMOGABA4  &  22/43/86/129  &  20    \\
2013-04-11  &  iMOGABA5  &  22/43/86/129  &  30    \\
2013-05-08  &  iMOGABA6  &  22/43/86  &  25    \\
2013-09-24  &  iMOGABA7  &  22/43  &  25    \\
2013-10-15  &  iMOGABA8  &  22/43/86  &  20    \\
2013-11-20  &  iMOGABA9  &  22/43/86/129  &  30    \\
2013-12-24  &  iMOGABA10  &  22/43/86/129  &  35    \\
2014-01-02  &  PAGaN  &  22/43  &  240    \\
2014-01-27  &  iMOGABA11  &  22/43  &  35    \\
2014-02-28  &  iMOGABA12  &  22/43/86/129  &  55    \\
2014-03-05  &  PAGaN  &  86  &  109    \\
2014-03-22  &  iMOGABA13  &  22/43/86/129  &  35    \\
2014-04-22  &  iMOGABA14  &  22/43/86/129  &  35    \\
2014-06-13  &  iMOGABA15  &  22/43  &  45    \\
2014-09-01  &  iMOGABA16  &  22/43/86  &  30    \\
2014-09-27  &  iMOGABA17  &  22/86  &  30    \\
2014-10-29  &  iMOGABA18  &  22/43/86  &  30    \\
2014-11-28  &  iMOGABA19  &  22/43/86  &  25    \\
2014-12-26  &  iMOGABA20  &  22/43/86  &  40    \\
2015-01-15  &  iMOGABA21  &  22/43/86  &  35    \\
2015-02-24  &  iMOGABA22  &  22/43/86  &  35    \\
2015-03-26  &  iMOGABA23  &  22/43/86/129  &  35    \\
2015-04-30  &  iMOGABA24  &  22/43/86/129  &  35    \\
2015-05-27  &  PAGaN  &  43/129  &  251    \\
2015-09-24  &  iMOGABA26  &  22/43/86/129  &  30    \\
2015-10-23  &  iMOGABA27  &  22/43/86  &  35    \\
2015-11-01  &  PAGaN  &  22/86  &  250    \\
2015-11-04  &  PAGaN  &  43  &  250    \\
2015-11-30  &  iMOGABA28  &  22/43/86/129  &  35    \\
2015-12-28  &  iMOGABA29  &  22/43/86/129  &  35    \\
2016-01-13  &  iMOGABA30  &  22/43/86/129  &  40    \\
2016-02-11  &  iMOGABA31  &  22/43/86/129  &  30    \\
2016-03-01  &  iMOGABA32  &  22/43/86/129  &  30    \\
\bottomrule
\end{tabular}
\tabnote{
   $^{\rm a}$  Frequencies with successful imaging.
\\ $^{\rm b}$  Total on-source time per session.
}
\end{table}

\section{Observations and Data Reduction \label{sec:2}}

We primarily use KVN VLBI data obtained within the interferometric MOnitoring of GAmma-ray Bright AGN (iMOGABA) program.\footnote{\url{http://radio.kasi.re.kr/sslee/}} iMOGABA monitors more than 30 $\gamma$-ray bright AGN via monthly (except during KVN maintenance times) KVN observations in single polarization (LCP) \citep{lee2016}. iMOGABA uses a snapshot mode, with several 5-minute observations spread over a few hours for each target, resulting in total integration times of few tens of minutes per source. The first observation (iMOGABA1) was performed in December 2012, with observations going on ever since. More details about iMOGABA can be found in \citet{lee2013, algaba2015, hodgson2016, lee2016}. As for the $\gamma$-ray activity of BL Lac, we follow the public database\footnote{\url{http://fermi.gsfc.nasa.gov/ssc/data/access/lat/msl_lc/}} (weekly light curve) of the \emph{Fermi} Large Area Telescope (\emph{Fermi}-LAT).

We collected data for BL Lac from 30 iMOGABA sessions (iMOGABA2 to iMOGABA32) from January 2013 (MJD 56308) to March 2016 (MJD 57448) for all four KVN frequencies. iMOGABA25 failed due to recording problems at Ulsan station. Due to technical or weather issues, we could not always obtain a full set (i.e., 22--129 GHz) of images in an iMOGABA session. Occasionally, the 129-GHz band had to be excluded due to instrumental problems. Some sessions suffered from an insufficient $uv$ coverage at some frequencies or missing baselines after data reduction; hence, we discarded those images. In order to increase the number of measurements over time and to perform a cross-check, we included five KVN observations executed independently by the Plasma-physics of Active Galactic Nuclei (PAGaN) project \citep{kim2015, oh2015}. PAGaN employs full-track dual-polarization observations of selected targets, typically reaching a few hours of integration time per source and per session. Due to different on-source times, the dynamic ranges of the PAGaN data at 22, 43, and 86 GHz are about 1.5--2 times higher than the corresponding iMOGABA data on average. At 129~GHz, only one PAGaN data set of poor quality was available; thus a meaningful comparison is not possible. We summarize the iMOGABA and PAGaN observations in Table~\ref{tab:obslog}.

We reduced our data with the software packages \texttt{AIPS}\footnote{Astronomical Image Processing System, distributed and maintained by the National Radio Astronomy Observatory of the U.S.A. (\url{http://www.aips.nrao.edu/index.shtml})} and \texttt{Difmap} \citep{shepherd1994}. We followed the standard \texttt{AIPS} steps for calibration of amplitudes and phases, bandpass, and opacity correction. To compensate for known amplitude losses, we multiplied a factor 1.1 to \texttt{APARM(1)} within the \texttt{AIPS} task \texttt{APCAL} (see \citealt{lee2015}). Detecting and imaging our target in data obtained at 86 and 129 GHz proved difficult because of high levels of visibility phase noise. Due to atmospheric and instrumental effects, such as tropospheric errors caused by the inhomogeneous distribution of water vapor, and high receiver temperatures (e.g., \citealt{rioja2011, vidal2012, algaba2015, rioja2015}), it is challenging to obtain good signal-to-noise (S/N) ratios, well-aligned phases, and sufficient coherence times at those high frequencies.

In order to increase coherence times, we applied the frequency phase transfer (FPT) technique \citep{rioja2011, algaba2015, zhao2015, hodgson2016} to parts of our data. Based on the assumption that the tropospheric path delay is independent of the observing frequency, we transferred phase solutions obtained for 43~GHz, scaled up by the ratio of frequencies, to the higher frequencies (86 and 129 GHz) by using the option \texttt{XFER} in the \texttt{AIPS} task \texttt{SNCOR}. We obtained improved coherence times of about five minutes on average (according to \citealt{algaba2015} and \citealt{rioja2015}, coherence times longer than five minutes can be reached even at 129 GHz) and thus improved S/N. Using FPT for iMOGABA data was suggested to us after we had started our analysis \citep{algaba2015}. Therefore, for iMOGABA2 to iMOGABA23, we applied FPT only to sessions with poor data quality at 86 and/or 129 GHz, and to all data after iMOGABA23. The difference in measured flux between data with and without FPT is within 5\% at 86 and 129 GHz, in agreement with the statistical errors -- meaning the application (or not) of FPT does not bias our flux measurements.

We imaged our radio data manually with \texttt{Difmap}, using \texttt{CLEAN} \citep{hogbom1974} deconvolution and phase self-calibration. (We could not apply amplitude self-calibration because the KVN consists of three antennas.) We derived positions, flux densities, and sizes of radio components by fitting circular Gaussian profiles to them. We estimated and propagated parameter errors as suggested by \citet{formalont1999} and \citet{lee2008}. Before imaging the $uv$ data, we flagged outlying visibilities (showing up as down-streaming features in visibility amplitude -- $uv$ radius diagrams) for a given antenna whenever they were clearly caused by bad antenna pointings. As far as possible, we compared our imaging results to those from the VLBA--Boston University Blazar Monitoring Program\footnote{\url{http://www.bu.edu/blazars/VLBAproject.html}} as additional quality check. Our images showed the presence of a radio component (the `knot') that corresponds to the jet. However, (re)detection of this component in the entire dataset proved difficult especially at the higher frequencies (cf. Figure~\ref{fig:2}).  Each iMOGABA session observes simultaneously at 22, 43, 86, and 129 GHz for 24 hours, with 6 scans (corresponding to 30 minutes) on average for BL~Lac, with only three antennas. The resulting sparse $uv$ coverage leads to conspicuous side lobes and beam patterns that are (largely) identical in all four bands (when taking into account scaling with frequency), thus implying a risk of confusing artifacts with the actual radio jet. Therefore, we focus on analyzing the VLBI core in this study.

\section{Results\label{sec:3}}

In this section, we present the multi-frequency behavior of BL Lac as obtained from our KVN data. During our observations, two $\gamma$-ray (0.1$\--$300 GeV) outbursts occurred in November 2013 and March 2015. The first outburst, as well as its optical counterpart, was reported by \citet{ramakrishnan2016} and \citet{gomez2016}. The second $\gamma$-ray outburst is visible in the data (their Figure~1) of \citet{sandrinelli2017}. Continuous data are provided by the weekly \emph{Fermi}-LAT light curve\footnote{\url{http://fermi.gsfc.nasa.gov/ssc/data/access/lat/msl_lc/source/BL_Lac}} (which is preliminary and used as a trigger for follow-up multi-wavelength observations). One can easily recognize the two outbursts in the \emph{Fermi}-LAT light curve, peaking at 5.3 $\times$ $10^{-7}$ photons cm$^{-2}$ s$^{-1}$ at around MJD 56620 and at 7.2 $\times$ $10^{-7}$ photons cm$^{-2}$ s$^{-1}$ at around MJD 57110, respectively. The $\gamma$-ray flux remained on a high level continuously after the second outburst; this is different from the first outburst which lasted $\sim$6 months.

\subsection{Radio Morphology of BL Lac\label{sec:3.1}}

\begin{figure}[t!]
\centering
\includegraphics[trim=3mm 0mm 20mm 10mm, clip, width=84mm]{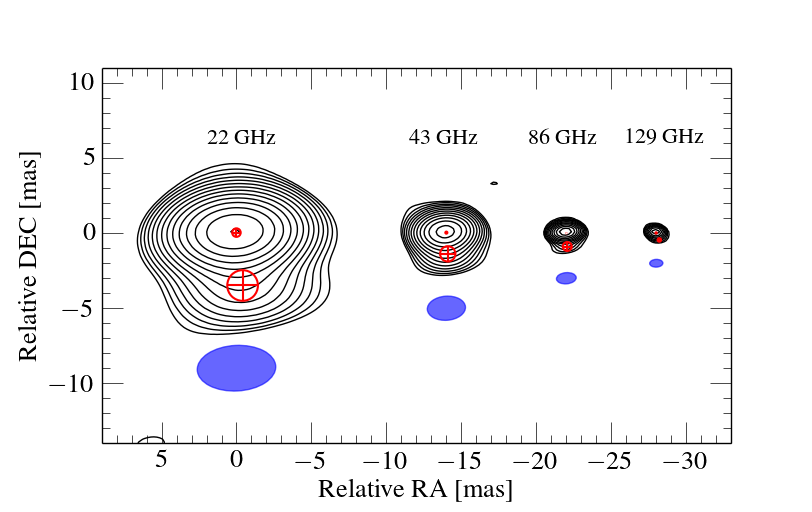}
\caption{KVN total intensity maps of BL Lac at 22, 43, 86, and 129 GHz, from observation iMOGABA23 (cf. Table~\ref{tab:obslog}). To avoid overlap, contours are offset in RA. Identified radio components are marked with red $\oplus$ symbols representing size and position as derived from fitting Gaussian circular profiles. Blue ellipses indicate the corresponding clean beams. Contour levels start at 1.1\%, 1.9\%, 2.41\%, and 6.9\% of the peak intensity at 22, 43, 86, and 129 GHz, respectively, and increase by factors $\sqrt{2}$.\label{fig:1}}
\end{figure}

In a typical KVN map, our observations show two radio components, the VLBI `core' and a single `knot' (in the standard nomenclature used by, e.g., \citealt{kadler2008}) south of the core that corresponds to the well-known jet of BL Lac (e.g., \citealt{stirling2003}). Figure~\ref{fig:1} shows a KVN map of BL Lac at 22, 43, 86, and 129 GHz (with contours offset in right ascension to avoid overlap); this map is typical for our iMOGABA data when regarding contours starting at three times the rms noise level. The KVN cores of BL~Lac at different frequencies (always at the map center) can be assumed to be identical; across our frequencies, core shift is less than about 0.2~mas \citep{dodson2017}. The KVN core light curves (see Figure~\ref{fig:2}) are very similar in their structure, further supporting the conclusion that we observed the same structure at all frequencies. The location of radio knots is inversely proportional to the observing frequencies, with knots being located at roughly 3.4~mas, 1.8~mas, 0.9~mas, and 0.6~mas from the core at 22~GHz, 43~GHz, 86~GHz, and 129~GHz respectively. We expect the jet components of BL~Lac at such distances to be optically thin (e.g., \citealt{denn2000}). Hence, we concluded that our observations of the jet of BL~Lac are heavily influenced by sensitivity and $uv$ coverage of KVN, forcing us to refrain from quantitative conclusions.

Qualitatively, we note that the jet appears longer at lower frequencies (cf. Figure~\ref{fig:1}). The same trend can also be found in \citet{osullivan2009} and \citet{abdo2011}. This is a frequently observed feature of blazar jets which suggests that emission at lower frequencies originates from more extended regions in blazar jets \citep{valtaoja1992}. In our case however, this observation is probably an artifact of the limited sensitivity of our KVN maps. The sensitivity decreases with increasing frequency, meaning that low-level flux contours are increasingly suppressed at higher frequencies.

\subsection{Radio Light Curves\label{sec:3.2}}

\begin{figure}[t!]
\centering
\includegraphics[trim=12mm 10mm 18mm 20mm, clip, width=84mm]{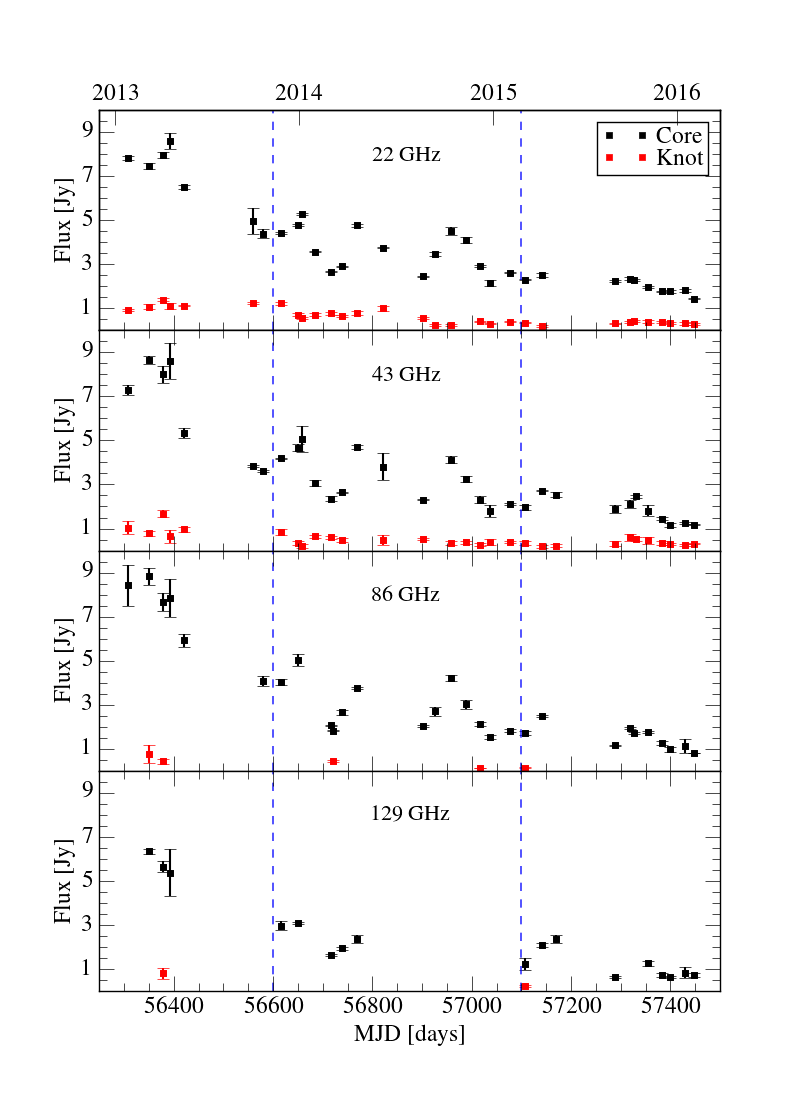}
\caption{22, 43, 86, and 129-GHz light curves of BL Lac obtained from KVN observations (both iMOGABA and PAGaN). The two blue dashed lines indicate $\gamma$-ray outbursts. Core (black) and knot (red) are plotted separately.\label{fig:2}}
\end{figure}

We monitored the flux densities of radio core and knot throughout our observing time as far as possible. Due to occasionally insufficient $uv$-coverage and/or high image noise, imaging of BL Lac or detection of the knot was not always possible at one or more frequencies. Especially, at 86 and 129 GHz we could detect a knot only in a few observations. Figure~\ref{fig:2} shows the KVN multi-frequency light curves of BL Lac at 22, 43, 86, and 129 GHz. The light curves for all four frequencies follow very similar trends. The light curves of the core show relatively high flux levels, up to about 9~Jy, at the beginning and decay over time, reaching levels around 1~Jy at the end of our observations. We note that our observations cover the decay phase of the 2012 radio outburst \citep{wehrle2016}. Flux densities of the knots are around 1~Jy at the beginning and decrease gradually by a factor of around two. We do not find obvious radio counterparts to the two $\gamma$-ray outbursts, even though any of the three local KVN core flux maxima located between the two $\gamma$-ray outbursts is a candidate for a counterpart to the first outburst.

\begin{figure}[t!]
\centering
\includegraphics[trim=10mm 3mm 12mm 12mm, clip, width=84mm]{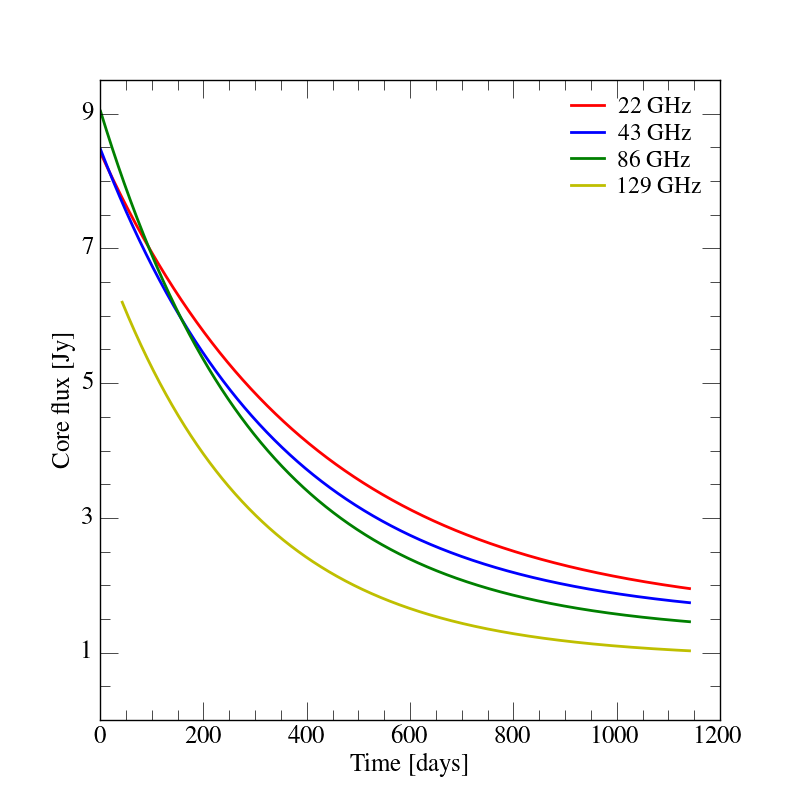}
\caption{Exponential fits to the light curves of the core, quantifying the long-term flux evolution. Different colors indicate the four observing frequencies.\label{fig:3}}
\end{figure}

To provide a quantitative description of the long-term flux evolution of the core, we fitted exponential functions
\begin{equation}
\label{eq:exponential}
S_{\nu} = A_{\nu} \exp(-B_{\nu}t) + C_{\nu}
\end{equation}
with $B_{\nu}\equiv1/\tau_{\nu}$ and $\tau_{\nu}$ being the decay time scale at frequency $\nu$, to each core light curve. Given that the intrinsic scatter of the data points about the long-term trend $\--$ AGN light curves show red-noise like variability patterns with smaller variability amplitudes at shorter time scales (\citealt{press1978}; recently, e.g., \citealt{park2016}) $\--$ is much larger than their statistical errors, we applied uniformly-weighted least-squares minimization. Figure~\ref{fig:3} illustrates the fitting results for the four light curves (omitting the actual data for clarity). At the very beginning of our observations, the fluxes at 22, 43, and 86 GHz are in good agreement; compared to these, the 129~GHz flux is lower by about 30\%. Due to this, the flux density measurements at 129~GHz may not be robust and extra care should be taken when interpreting them and associated quantities. Later, fluxes are consistently higher at lower frequencies.

\begin{figure}[t!]
\centering
\includegraphics[trim=12mm 2mm 18mm 15mm, clip, width=84mm]{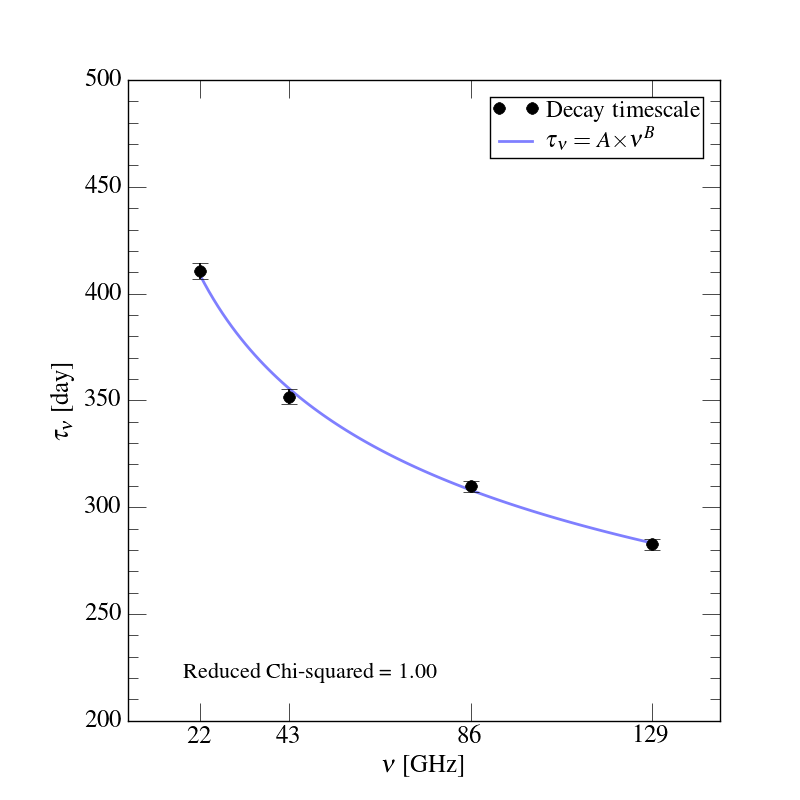}
\caption{Decay timescale of the core flux as function of observing frequency. Both axes are linear. Errors have been re-scaled such that $\chi^2/{\rm d.o.f.} = 1$. The blue solid line represents the best power-law fit, with a power-law index of $-0.208\pm0.007$.\label{fig:4}}
\end{figure}

We obtain a decay timescale for each light curve from the exponential model. The timescales are $411\pm85$ days at 22 GHz, $352\pm79$ days at 43 GHz, $310\pm57$ days at 86 GHz, and $283\pm55$ days at 129 GHz. The formal errors provided by the fits reflect the intrinsic short-term variability of the source fluxes rather than the actual statistical uncertainties. Taking this into account, we note that the decay timescale depends on the observing frequency. We fit a power-law to the timescale $\--$ frequency data and re-scale the error bars such that $\chi^2/{\rm d.o.f.} = 1$. Figure~\ref{fig:4} shows the resulting relation. The re-scaled errors on the timescales are $\pm$3.9 days at 22 GHz, $\pm$3.6 days at 43 GHz, $\pm$2.6 days at 86 GHz, and $\pm$2.5 days at 129 GHz. The best-fit line returns a power-law index of $-0.208\pm0.007$. Using a linear approximation for the frequency range covered by our data (and applying, again, error re-scaling), we find a slope of $-1.05\pm0.23$ day/GHz -- in other words, for each additional GHz in frequency, the decay timescale drops by about one day in the frequency range 22$\--$129 GHz.

\subsection{Spectral Indices and Spectrum of the Core\label{sec:3.3}}


We analyzed the evolution of the spectrum of the core in view of a possible evolution of its opacity, beginning with spectral indices from pairs of adjacent frequencies (`pairwise indices'). As mentioned in Section~\ref{sec:2}, we could not always obtain measurements for all four frequencies, leading to different numbers of pairwise indices for the three frequency pairs. Due to the small time gap of three days between them, we treated the PAGaN observations of MJD 57327 and 57330 (2015 November 1 and 4) as a single measurement. We defined the spectral index $\alpha$ via $S_\nu \propto \nu^{\alpha}$, where $S_\nu$ is flux density of the core and $\nu$ is observing frequency. We adopt the conventional criterion that spectra with $\alpha < -0.5$ and $\alpha > -0.5$ are `steep' and `flat', respectively \citep[e.g.,][]{orienti2007, netzer2013}. Figure~\ref{fig:5} shows the pairwise spectral indices of the core as function of time. We obtained 31 spectral index values with an average value of $-0.17$ at 22$\--$43~GHz, 27 indices with an average value of $-0.14$ at 43$\--$86~GHz, and 16 indices with an average value of $-0.89$ at 86$\--$129~GHz. Our observations show for the first time explicitly that the VLBI core of BL~Lac is optically thick at frequencies higher than 43~GHz over several years. This agrees with expectations that the core should be optically thick because of synchrotron self-absorption (SSA) \citep{denn2000, wehrle2016}. The spectral indices for the range from 22 to 86 GHz slightly decrease with time, in agreement with the trend illustrated by Figure~\ref{fig:3}. Figure~\ref{fig:5} shows indication for a steady spectral hardening at 43$\--$86~GHz from $\alpha\approx-0.3$ to $\alpha\approx-0.1$ for a few months around the time of the second $\gamma$-ray event, just after a rapid spectral softening. For the frequency pair 86$\--$129~GHz, we see indication for the core becoming optically thin at these high frequencies; given the large errors, this conclusion has to be handled with care however.

Given that KVN provides multi-frequency data sets, we can obtain mm-wavelength spectra that are more sophisticated than the use of pairwise spectral indices. Whenever possible, we fit a power-law to spectra of the core consisting of the flux data that were obtained simultaneously at 22, 43, 86, and 129~GHz, thus making the standard assumption that BL Lac shows a synchrotron spectrum like all radio-loud AGN \citep[e.g.,][]{rani2011, trippe2011}. Figure~\ref{fig:6} illustrates the estimation of a spectrum and shows the evolution of the spectral index $\alpha$ as function of time. We obtained 16 spectra in total. As expected from the pairwise spectral indices, BL~Lac's core spectrum is flat (within errors) at the beginning and steepens slightly throughout our observations, down to $\alpha\approx-0.5$. We summarize all spectral indices in Table~\ref{tab:spinx}.

\begin{table}[t!]
\caption{Spectral indices of the core as function of time\label{tab:spinx}}
\centering
\setlength{\tabcolsep}{2.5pt}
\begin{tabular}{lrrrr}
\toprule 
MJD$^{\rm a}$ & $\alpha_{22-43}$ ~~ & $\alpha_{43-86}$ ~~ & $\alpha_{86-129}$  ~ & $\alpha_{22-129}^{\rm c}$ ~ \\
\midrule
56308  &  $-$0.11$\pm$0.05  &  0.21$\pm$0.17  &    &      \\
56350  &  0.22$\pm$0.04  &  0.03$\pm$0.07  &  $-$0.83$\pm$0.12  &  $-$0.04$\pm$0.13    \\
56379  &  0.01$\pm$0.07  &  $-$0.06$\pm$0.10  &  $-$0.76$\pm$0.17  &  $-$0.14$\pm$0.09    \\
56393  &  0.00$\pm$0.15  &  $-$0.13$\pm$0.21  &  $-$0.96$\pm$0.57  &  $-$0.20$\pm$0.11    \\
56420  &  $-$0.28$\pm$0.06  &  0.15$\pm$0.09  &    &      \\
56559  &  $-$0.36$\pm$0.17  &    &    &      \\
56580  &  $-$0.28$\pm$0.07  &  0.17$\pm$0.08  &    &      \\
56616  &  $-$0.08$\pm$0.02  &  $-$0.05$\pm$0.05  &  $-$0.77$\pm$0.19  &  $-$0.17$\pm$0.08    \\
56650  &  $-$0.03$\pm$0.05  &  0.11$\pm$0.09  &  $-$1.22$\pm$0.14  &  $-$0.14$\pm$0.15    \\
56659  &  $-$0.06$\pm$0.17  &    &    &      \\
56684  &  $-$0.21$\pm$0.06  &    &    &      \\
56716  &  $-$0.18$\pm$0.08  &  $-$0.19$\pm$0.08  &  $-$0.57$\pm$0.10  &  $-$0.24$\pm$0.04    \\
56721  &    &    &    &      \\
56738  &  $-$0.14$\pm$0.02  &  0.01$\pm$0.06  &  $-$0.77$\pm$0.13  &  $-$0.17$\pm$0.08    \\
56769  &  $-$0.02$\pm$0.03  &  $-$0.32$\pm$0.03  &  $-$1.15$\pm$0.20  &  $-$0.30$\pm$0.13    \\
56821  &  0.02$\pm$0.23  &    &    &     \\
56901  &  $-$0.08$\pm$0.03  &  $-$0.19$\pm$0.05  &    &      \\
56927  &    &    &    &      \\
56959  &  $-$0.12$\pm$0.08  &  0.03$\pm$0.07  &    &      \\
56989  &  $-$0.33$\pm$0.08  &  $-$0.10$\pm$0.11  &    &      \\
57017  &  $-$0.33$\pm$0.10  &  $-$0.12$\pm$0.12  &    &      \\
57037  &  $-$0.26$\pm$0.25  &  $-$0.24$\pm$0.25  &    &      \\
57077  &  $-$0.31$\pm$0.03  &  $-$0.21$\pm$0.07  &    &      \\
57107  &  $-$0.23$\pm$0.08  &  $-$0.20$\pm$0.11  &  $-$0.83$\pm$0.55  &  $-$0.30$\pm$0.06    \\
57142  &  0.11$\pm$0.06  &  $-$0.12$\pm$0.03  &  $-$0.44$\pm$0.11  &  $-$0.09$\pm$0.07    \\
57169  &    &    &    &     \\
57289  &  $-$0.24$\pm$0.14  &  $-$0.69$\pm$0.14  &  $-$1.57$\pm$0.19  &  $-$0.56$\pm$0.15    \\
57318  &  $-$0.15$\pm$0.14  &  $-$0.13$\pm$0.14  &    &      \\
57327  &  0.10$\pm$0.05  &  $-$0.52$\pm$0.05  &    &      \\
57330$^{\rm b}$  &  0.10$\pm$0.05  &  $-$0.52$\pm$0.05  &    &      \\
57356  &  $-$0.10$\pm$0.21  &  $-$0.04$\pm$0.21  &  $-$0.85$\pm$0.28  &  $-$0.19$\pm$0.09    \\
57384  &  $-$0.29$\pm$0.06  &  $-$0.18$\pm$0.12  &  $-$1.45$\pm$0.33  &  $-$0.38$\pm$0.12    \\
57400  &  $-$0.62$\pm$0.13  &  $-$0.23$\pm$0.21  &  $-$1.04$\pm$0.36  &  $-$0.51$\pm$0.08    \\
57429  &  $-$0.56$\pm$0.07  &  $-$0.11$\pm$0.40  &  $-$0.76$\pm$1.03  &  $-$0.40$\pm$0.07    \\
57448  &  $-$0.29$\pm$0.05  &  $-$0.53$\pm$0.05  &  $-$0.24$\pm$0.26  &  $-$0.39$\pm$0.04    \\
\bottomrule
\end{tabular}
\tabnote{
   $^{\rm a}$  Same dates as in Table~\ref{tab:obslog} in units of MJD.
\\ $^{\rm b}$  Quasi-simultaneous with the data of MJD 57327.
\\ $^{\rm c}$  Spectral index using 22, 43, 86, and 129-GHz data.
}
\end{table}

\section{Discussion\label{sec:4}}

The behavior of multi-waveband light curves provides important clues on the physical processes within blazar jets. Even though opacity and flux evolution depend on frequency, correlations between the flux variability at $\gamma$-rays and that at lower frequencies have been observed in blazar jets (as noted in Section~\ref{sec:1}). Unfortunately, there is no clear picture as yet: observed correlations (or the absence thereof) seem to change over time and from source to source \citep[e.g.,][]{marscher2014}. Simultaneous multi-frequency  observations by KVN may shed light on the connection between $\gamma$-ray emission and emission at high radio frequencies (up to $\sim$130~GHz in our case). In this Section, we discuss the long-term decay of the 2012 radio outburst, the nature of the KVN core, and the connection between the two $\gamma$-ray outbursts and radio jet.

\subsection{Variability and Cooling Time Scales\label{sec:4.1}}

We find the highest fluxes for BL~Lac right at the beginning of our observations. \citet{wehrle2016} reported a strong outburst of BL~Lac at 1.3 mm (225 GHz) at the end of 2012 preceded by strong $\gamma$-ray flaring. Accordingly, we may identify the 2012 radio outburst with the radio counterpart of the $\gamma$-ray flare(s) of the same year. Arguably, our observations cover the decay phase of the 2012 radio flare \citep[see also][]{gaur2015}. An exponential decay of radio luminosity is a common characteristic of blazar flares \citep{valtaoja1999, tavares2011, trippe2011, chatterjee2012, marscher2014, guo2016, park2016}. Unfortunately, our observations do not cover the rising stage of the radio outburst. Typically, the decay timescale of a flare in a radio AGN is about 1.3 times longer than its rise timescale \citep{valtaoja1999}. This suggests rise timescales of $\sim$316 days at 22 GHz, $\sim$271 days at 43 GHz, $\sim$238 days at 86 GHz, and $\sim$218 days at 129 GHz.

\begin{figure}[t!]
\centering
\includegraphics[trim=10mm 5mm 12mm 14mm, clip, width=84mm]{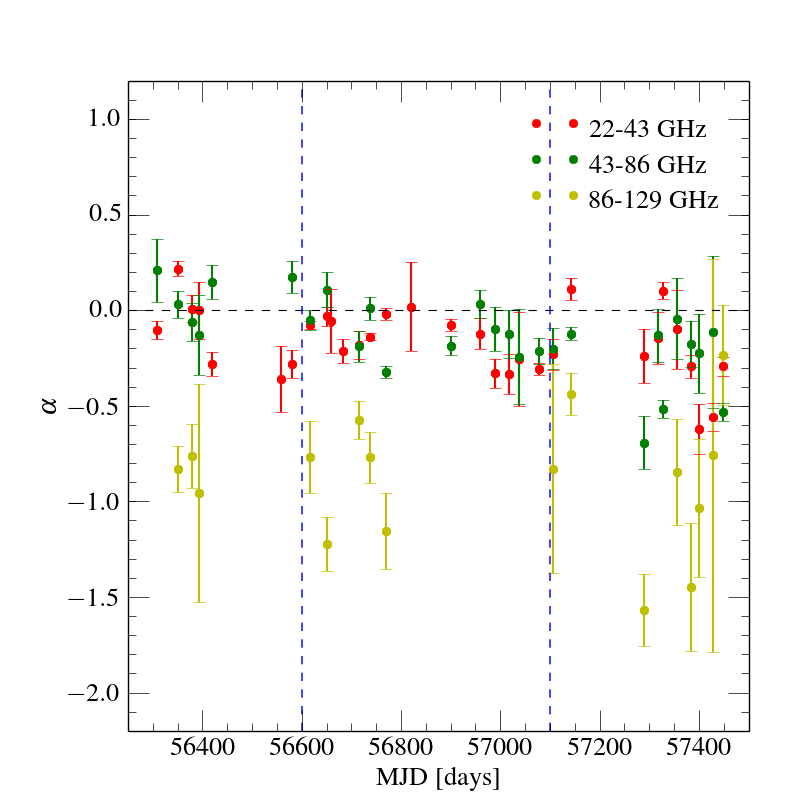}
\caption{Pairwise spectral indices of the core as function of time, for the frequency pairs 22--43~GHz, 43--86~GHz, and 86--129~GHz. Different colors correspond to different frequency pairs. Vertical blue dashed lines mark the occurrence times of the two $\gamma$-ray outbursts.\label{fig:5}}
\end{figure}

\begin{figure}[t!]
\centering
\includegraphics[trim=2mm 0.7mm 12mm 10mm, clip, width=84mm]{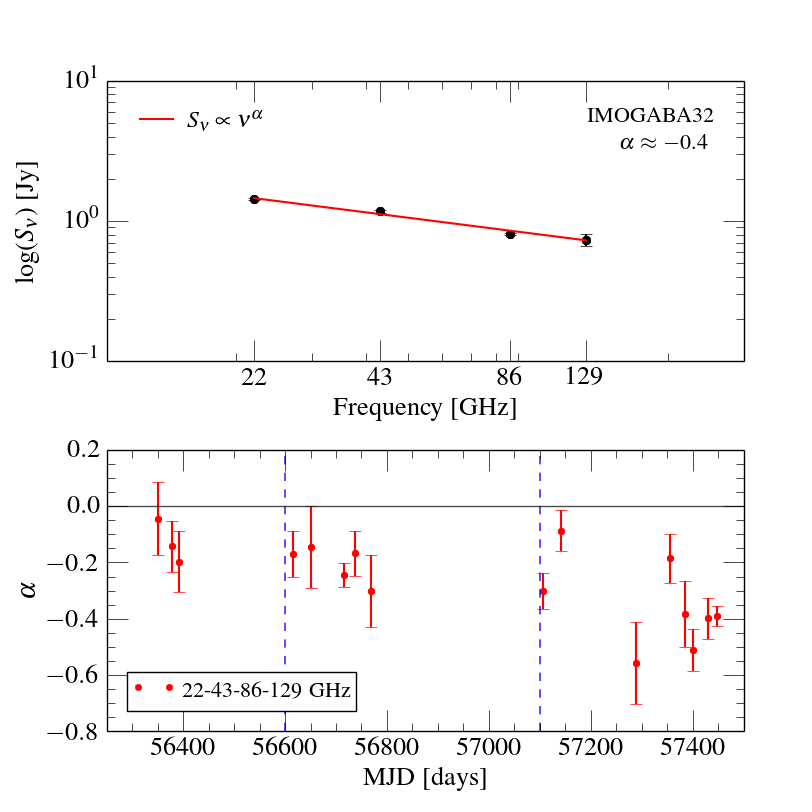}
\caption{Radio spectra of the core from power-law fits to 22, 43, 86, and 129-GHz flux values. \emph{Top:} An example spectrum, from data set iMOGABA32. \emph{Bottom:} Spectral index $\alpha$ as function of time. Vertical blue dashed lines mark the occurrence times of the two $\gamma$-ray outbursts.\label{fig:6}}
\end{figure}

The radio core of blazar jets is usually a compact, unresolved, and flat spectrum source that emits synchrotron radiation with high power. The radio core is known to be located a few parsecs downstream from the central engine \citep{tavares2011}. As noted by \citet{spada2001}, synchrotron radiation is the main cooling mechanism for relativistic electrons in blazar jets at such distances from the central engine, while IC dominates inside the broad line region ($\sim$10$^{15}$~m). Here, we assume synchrotron cooling to be the dominant cooling mechanism in the core region \citep{sokolov2004}. (In general, inverse Compton losses are less important than synchrotron losses in a stationary synchrotron source; \citealt{pachol1970}). A characteristic synchrotron cooling time can be defined via $\tau_{\rm cool} = E/\dot{E}$, with the electron energy $E = \gamma m c^2$ (with Lorentz factor $\gamma$, electron mass $m$, and speed of light $c$) and $\dot{E}=\mathrm{d}E/\mathrm{d}t$ (with time $t$) being the power radiated by a gyrating electron \citep[e.g.,][]{rybicki1997}. In SI units and after transforming to the observer frame, the cooling time is 
\begin{equation}
\label{eq:cool}
\tau_{\rm cool} = 7.74\,\left[\frac{\delta}{1+z}\right]^{-1}B^{-2}\,\gamma^{-1} ~ {\rm seconds}
\end{equation}
where $B$ is the magnetic field strength, $\delta$ is the Doppler factor, and $z$ is the redshift. For BL~Lac, $z=0.069$ and $\delta\approx7$ \citep{hovatta2009}. Cooling times on the order of one year (assuming $\tau_{\rm cool}\approx\tau_{\nu}$; see Section~\ref{sec:3.2}) can be translated to $B\sim2\,\mu$T and $\gamma\sim10\,000$ which is consistent with the typical blazar magnetic field of $B \sim$ 1 $\--$ 10 $\mu$T \citep{lewis2016}. Moreover, \citet{agarwal2017} obtained the magnetic field of the BL~Lac core at 4.8, 8, 14.5, and 22.2 GHz. They found $B$ to increase with frequency, with $B\sim1\,\mu$T at 22~GHz on average. Electron Lorentz factors $\gamma\sim10\,000$ are consistent with the fact that the peak frequencies of the synchrotron and IC humps in BL~Lac's SED differ by about eight orders of magnitude (cf. Figure 27 of \citealt{wehrle2016}).

The presence of longer decay time scales at lower frequencies, and the subsequent stratification of flux densities as function of frequency, might be an effect of longer synchrotron cooling times at lower electron energies \citep{marscher2008}. In general, the synchrotron cooling time is longer at lower frequencies \citep{rybicki1997}, which can be understood from the basic scaling relations of synchrotron radiation. An electron gyrating about a magnetic field line with angular frequency $\omega_{\rm B}$ emits synchrotron radiation with a critical frequency $\nu_{\rm c}\propto\gamma^3\omega_{\rm B}$ \citep[e.g.,][]{rybicki1997}. With $\omega_{\rm B}\propto\gamma^{-1}$, $\nu_{\rm c}\propto\gamma^2$. Because $\tau_{\rm cool}\propto\gamma^{-1}$ (Equation~\ref{eq:cool}), the cooling time follows $\tau_{\rm cool}\propto\nu_{\rm c}^{-1/2}$ for \emph{optically thin} plasmas. In addition, one needs to consider that electron energies commonly follow power-law distributions of the form $N(\gamma) \propto \gamma^{p}$ (with $p<0$); as the cooling time is a function of $\gamma$, high-energy electrons will cool down faster. Accordingly, single-power laws are widely used to describe synchrotron losses in blazar jets \citep[see also][]{bottcher2003, sokolov2004}.

\begin{figure}[t!]
\centering
\includegraphics[trim=14mm 0mm 27mm 10mm, clip, width=84mm]{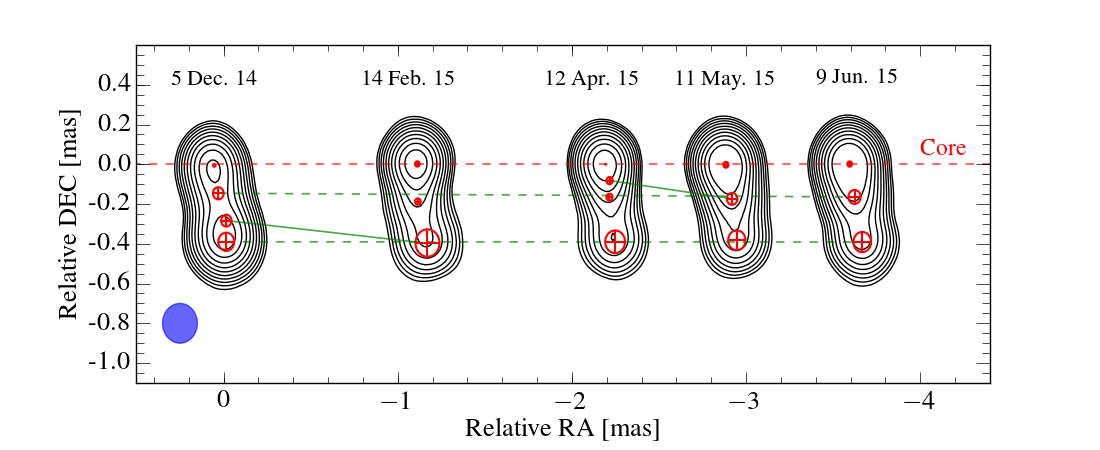}
\caption{43-GHz VLBA maps (from the BU data base) of BL Lac obtained around the time of the March 2015 $\gamma$-ray outburst. Contours represent total intensity levels ranging from 4\% to 90.5\% of the peak intensity. Red $\oplus$ denote circular Gaussian components, the blue ellipse indicates the clean beam. Maps are shifted along the abscissa for clarity; all maps are restored with the same $0.2\times0.2$~mas beam.\label{fig:8}}
\end{figure}

In our case however, the KVN core of BL~Lac is \emph{optically thick} (cf. Section~\ref{sec:3.3}), meaning that observations at different frequencies probe different regions with different magnetic field strengths. In general, $B \propto d^{-1}$, with $d$ being the distance from the central engine, in the VLBI core \citep{marscher1995, lobanov1998}. \citet{marscher1995} suggested that the observed $\gamma$ becomes the maximum Lorentz factor for the given electron distribution, $\gamma_{\rm max}$, at the VLBI core. We assume $\gamma_{\rm max}$ to be constant in the core region (e.g., \citealt{kaiser2006}). Our data (Section~\ref{sec:3.2}) follow a relatively shallow scaling relation $\tau\propto\nu^{-0.2}$. For a conical \citet{blandford1979} jet, different observing frequencies probe different plasma surfaces. This results in the well-known `core shift' relation $d \propto \nu^{-1}$ (\citealt{lobanov1998}; but see also \citealt{agarwal2017}). Such a scaling law is in tension with our observations: even for (approximately) constant values of $\gamma$, we would expect $B^2\gamma \propto d^{-2} \propto \nu^{2}$ and thus $\tau\propto\nu^{-2}$ -- which is much steeper than our observed relation. 

An alternative explanation for our result is provided by the assumption that the millimeter radio core of BL~Lac is a standing recollimation shock \citep{gomez1997, marscher2008}. Such a shock forms where the gas pressure in the jet does not match the gas pressure of the ambient interstellar medium \citep[e.g.,][]{cawthorne2013}. Numerical simulations by \citet{dodson2017} suggest that the radio core of BL~Lac follows the usual core shift relation at centimeter wavelengths and coincides with a recollimation shock (which is optically thin) at millimeter wavelengths \citep[see also][]{marscher2012, marti2016}. In this picture, the recollimation shock is a physical structure at a fixed position; high frequency (mm to sub-mm) radiation may originate entirely from the shock. The core shift scaling starts to deviate from the relation $d \propto \nu^{-1}$  at wavelengths shorter than about one centimeter. At sufficiently high observing frequencies, the synchrotron radiation originates from the same plasma surface at all frequencies; the scaling of magnetic field strength with frequency breaks down, i.e., $B^2\gamma$ and $\tau$ become independent of frequency. We suspect that the KVN frequency bands cover the frequency range where this transition occurs in BL~Lac, probably between 86 and 129~GHz. This interpretation is supported by the astrometric observations of \citet{dodson2017}, who find that the core shift relation beyond $\sim$30 GHz is much shallower than $d\propto\nu^{-1}$ $\--$ in agreement with the signature of a recollimation shock \citep[see also][]{wehrle2016, gomez2016}. \citet{osullivan2009} find the millimeter radio core of BL Lac to be located $\lesssim$0.5 pc from the jet base.

\subsection{Shock Evolution in the Core Region\label{sec:4.2}}

The energy distribution of electrons in relativistic AGN jets is commonly assumed to follow a power law \citep{hughes1985, ghisellini2002, maselli2010, rani2011, trippe2011, giommi2012, karamanavis2016}, resulting in power law synchrotron spectra $\--$ in agreement with our KVN data. Throughout our observations, the long-term evolution of the core spectrum follows the overall long-term evolution of the core light curves (cf. Figures~\ref{fig:3} and \ref{fig:6}). At the beginning, the spectrum is approximately flat ($\alpha\approx0$) while the core flux is at its highest levels; over time, the flux decreases and the spectrum becomes steeper. A hardening of the spectrum at higher fluxes is a signature of powerful shocks and is consistent with the known flux evolution of BL~Lac since the 2012 radio outburst \citep{ghisellini2002, gaur2015, wehrle2016}. The subsequent behavior of the spectrum (i.e., becoming steeper) over time indicates again that the source underwent radiative energy losses by synchrotron radiation and Compton scattering \citep{blumenthal1970}. A physical explanation is provided by the `generalized shock model' of \citet{valtaoja1992} which links the evolution of a shock to the observable evolution of its turnover peak frequency. As the shock evolves with time, the turnover peak frequency of the spectrum moves to lower frequencies $\--$ in good agreement with our observations (Figure~\ref{fig:6}).

The KVN core is consistent with being optically thick at least up to 86 GHz (cf. Figures~\ref{fig:5} and \ref{fig:6}) throughout our observations. This is in contrast to the observation that the core becomes optically thin at mm-wavelengths occasionally due to the emission from newly ejected jet components \citep{hodgson_thesis}. This apparent conflict can probably be resolved by taking into account that the KVN core is a blend of multiple components that would be spatially resolved by higher-resolution observatories \citep[see also][]{rioja2014, rioja2015, zhao2015}. The long-term steepening of the spectrum which we observe (Figure~\ref{fig:6}) might be caused by the jet emission becoming brighter relative to the core emission over time. This implies that the core spectrum might continue its steepening until a jet component is ejected, and then return to $\alpha\approx0$. At 129 GHz, the KVN core may be optically thin permanently -- as expected at sufficiently high frequencies. However, we cannot ignore possible instrumental effects of the KVN that can result in unwanted amplitude drops and which are supposed to be more severe at higher frequencies \citep{algaba2015, lee2016}.

\subsection{The Radio -- $\gamma$-ray Connection \label{sec:4.3}}

Radio and $\gamma$-ray light curves of blazars tend to be connected such that a $\gamma$-ray outburst precedes a radio outburst \citep{valtaoja1995, jorstad2001b, marscher2008, marscher2011, tavares2011, arlen2013, raiteri2013, moerbeck2014, ramakrishnan2015, karamanavis2016, wehrle2016}. It has been suggested that a moving disturbance originating near the jet base produces flares both at radio and $\gamma$-rays as it propagates along the jet flow \citep{marscher2008, moerbeck2014}. In this scenario, the $\gamma$-ray outburst is assumed to occur in the inner jet region (upstream of the radio core) where the jet is opaque at radio wavelengths; the radio outburst becomes observable once the disturbance has passed through the radio core region. To add complexity to this issue, observations indicate that the production site of $\gamma$-ray outbursts is located parsecs downstream of the radio core at least occasionally (e.g., \citealt{tavares2011}). In addition, recent studies (e.g., \citealt{wehrle2016}, \citealt{hodgson2017}) found recollimation shocks in blazar jets within 0.5 mas downstream from the core, and suggested the recollimation shocks as a candidate of the origin of $\gamma$-ray outbursts in blazar jets.

The typical rise time of radio outbursts is on the order of one year \citep[e.g.,][]{hovatta2008}. In blazars, the observed delays between $\gamma$-ray and radio outbursts range from hours to a few hundred days, with the $\gamma$-ray emission leading the radio emission \citep{arlen2013, jorstad2013, raiteri2013, moerbeck2014, ramakrishnan2015, wehrle2016}. Accordingly, the identification of the radio counterparts to a specific $\gamma$-ray outburst is not straightforward if significant variability such as the multiple outbursts (the three minor radio flux peaks) after the first $\gamma$-ray event in Figure~\ref{fig:2} is present in the radio light curve. In our case, we do not find obvious radio counterparts to the two $\gamma$-ray outbursts.

The apparent absence of a radio -- $\gamma$-ray correlation (like the absence of radio counterparts) in blazar jets or BL Lac objects has been noted before \citep{denn2000, tavares2011, orienti2013, moerbeck2014, ramakrishnan2016}. The lack of radio counterparts can result from radio and $\gamma$ emission orignating from different regions (e.g., \citealt{tavares2011}); in this scenario, $\gamma$ emission results from IC scattering near the core with seed photons from the accretion disk, the outflowing broad-line region (BLR), or the dusty torus. A complex variability pattern could cause a weak radio counterpart to be `buried' under radio outbursts that are unrelated to the $\gamma$-ray emission. Indeed, already \citet{valtaoja1999} noted that light curves of BL~Lac are difficult to decompose into individual outbursts. \citet{gaur2015} suggested geometry effects to be important, specifically variable Doppler boosting of the jet emission due to viewing angle changes. This might cause radio counterparts to be missed or underestimated.

Last but not least, the $\gamma$-ray outbursts of 2013 and 2015  might be `orphan' events that show no correlation with the luminosity at longer wavelengths. An orphan $\gamma$-ray outburst was observed in the blazar PKS 1510-089 by \citet{marscher2010}; \citet{macdonald2015} explained this outburst as the result of a disturbance passing through the inner jet region. To further complicate matters, BL~Lac shows (apparent) radio counterparts to $\gamma$-ray outbursts at least in some cases \citep{arlen2013, wehrle2016}. Not only the light curves but also the evolution of the radio spectrum can be related to the $\gamma$-ray activity of blazars. For the quasar 3C~454.3, \citet{jorstad2013} found that $\gamma$-ray outbursts coincide with increases in -- spectroscopically determined -- core opacity. As shown in Figure~\ref{fig:5}, the spectral index of BL~Lac's core at 43--86~GHz reaches a local minimum, with $\alpha\approx-0.3$, roughly two months before the second $\gamma$-ray outburst. From then on, the spectral index, and thus the core opacity, increases up to $\alpha\approx-0.1$ a few weeks after the $\gamma$-ray outburst. This can be explained by the passage of a newly ejected jet component through the core: while in the core region, the new component is obscured and temporarily increases (or dampens the long-term decrease of) the flux from the optically thick core, resulting in increased values of $\alpha$. This interpretation is strengthened by the results of \citet{arlen2013} who found a $\gamma$-ray outburst in June 2011 (at MJD 55711) to coincide with the ejection of a new jet component.

The connection between $\gamma$-ray outbursts and ejections of new components is a known feature of blazars \citep{marscher2010, marscher2011, arlen2013, jorstad2013, gomez2016, wehrle2016} which is, however, not a universally valid relation \citep{denn2000, jorstad2001b}. We used archival VLBA data from the BU data base to search for a new jet component in spring 2015; the maps are shown in Figure~\ref{fig:8}. Indeed, a new jet component might have appeared about 0.1 mas south of the VLBA core in April 2015 which, however, only shows up in the April 2015 map -- meaning we cannot claim the occurence of a new jet component with sufficient certainty.
If, indeed, a new jet component appeared in April 2015, this could explain the occurrence of an orphan $\gamma$-ray outburst in March 2015 via the `ring of fire' mechanism \citep{macdonald2015}. In this picture, an orphan $\gamma$-ray outburst arises from a relativistic plasma cloud moving through the inner jet region and precedes the emergence of a new radio jet component -- by $\sim$20 days in the case of PKS 1510-089, and possibly by a similar amount of time in case of BL~Lac in 2015 ($\sim$25 days).

\section{Summary\label{sec:5}}

We analyzed multi-frequency KVN radio data of BL Lac spanning three years, from January 2013 to March 2016 to study the connection between radio and $\gamma$-ray emission in blazar jets. We obtained light curves and spectra of the radio core at mm-wavelengths during a period when BL~Lac showed two $\gamma$-ray outbursts that occurred in November 2013 and March 2015. We summarize our main results in the following.
\begin{enumerate}

\item Our observations cover the aftermath of the major radio flare that occurred in 2012. During the entire time of our observations, the radio flux from the KVN radio core decayed exponentially, and two $\gamma$-ray outbursts occured in BL~Lac. We found decay time scales $\tau$ of 411$\pm$85 days (22~GHz), 352$\pm$79 days (43~GHz), 310$\pm$57 days (86~GHz), and 283$\pm$55 days (129~GHz), with formal errors including intrinsic short-term variability. Assuming synchrotron cooling, such decay times are consistent with magnetic fields $B\sim2\,\mu$T and electron Lorentz factors $\gamma\sim10\,000$.

\item We find that the flux decay time scales with observing frequency like $\tau\propto\nu^{-0.2}$. This scaling law is much shallower than the synchrotron energy loss expected from optical depth effects in a \citet{blandford1979} jet. Therefore, we suspect the KVN radio core to be a standing recollimation shock (e.g., \citealt{marscher2008}) -- in agreement with independent recent astrometric observations \citep{dodson2017} of the core shift.

\item The radio core spectrum is initially flat ($\alpha\approx0$) and steepens with decreasing flux, reaching $\alpha\approx-0.5$ at the end of our observations. This is in agreement with the `generalized shock model' of \citet{valtaoja1992} wherein the turnover peak moves to lower frequencies when the shock decays, with the shock corresponding to the 2012 radio flare.

\item There is no obvious radio counterpart to any of the two $\gamma$-ray outbursts covered by our study. Possible explanations are: (i) the $\gamma$-ray emission originates downstream of the core; (ii) the radio counterparts are masked by the complex intrinsic red-noise type variability; (iii) geometry effects like variations in Doppler boosting; (iv) there is actually no radio counterpart -- i.e., the $\gamma$-ray outbursts are `orphans'.

\item The temporary increase in the core opacity two months before the second $\gamma$-ray outburst, as well as the outburst itself, might both be due to the emergence of a new jet component. The detection of a new VLBA radio knot in April 2015 tentatively suggests that the second $\gamma$-ray event might be an orphan $\gamma$-ray event powered by the `ring of fire' mechanism.

\end{enumerate}
Overall, our results strengthen the case for a causal connection between the radio and $\gamma$-ray activity in BL~Lac, and potentially in blazars in general. Even though, we are still missing many  important details. Future studies will need to explore if recollimation shocks are a general features of blazar jets; under which circumstances one may expect radio-flux counterparts to $\gamma$-ray outbursts; and which physical mechanism causes orphan $\gamma$-ray flares.


\acknowledgments
We are grateful to the staff of the KVN who helped to operate the array and to correlate the data. The KVN and a high-performance computing cluster are facilities operated by the KASI (Korea Astronomy and Space Science Institute). The KVN observations and correlations are supported through the high-speed network connections among the KVN sites provided by the KREONET (Korea Research Environment Open NETwork), which is managed and operated by the KISTI (Korea Institute of Science and Technology Information). We also thank Vassilis Karamanavis (MPIfR Bonn) for valuable comments. We made use of 43 GHz VLBA data from the VLBA--Boston University Blazar Monitoring Program (VLBA-BU-BLAZAR), funded by NASA through the Fermi Guest Investigator Program. We acknowledge financial support by the National Research Foundation of Korea (NRF) via research grants 2015-R1D1A1A01056807 (D.K., S.T., J.O., T.L.), 2016-R1C1B2006697 (S.-S.L., S.K.), 2015-H1D3A1066561 (G.-Y.Z.), and 2014-H1A2A1018695 (J.P.).



\begin{thebibliography}{}


\bibitem[Abdo et al.(2011)]{abdo2011}
Abdo, A. A., et al. 2011, The First Fermi Multifrequency Campaign on BL Lacertae: Characterizing the Low-Activity State of the Eponymous Blazar, ApJ, 730, 101

\bibitem[Agarwal et al.(2017)]{agarwal2017}
Agarwal, A., et al. 2017, Core Shift Effect in Blazars, MNRAS, 469, 813

\bibitem[Albert et al.(2007)]{albert2007}
Albert, J., et al. 2007, Discovery of Very High Energy $\gamma$-ray Emission from the Low-Frequency-Peaked BL Lacertae Object BL Lacertae, ApJ, 666, L17

\bibitem[Algaba et al.(2015)]{algaba2015}
Algaba, J.-C., et al. 2015, Interferometric Monitoring of Gamma-ray Bright Active Galactic Nuclei II: Frequency Phase Transfer, JKAS, 48, 237

\bibitem[Arlen et al.(2013)]{arlen2013}
Arlen, T., et al. 2013, Rapid TeV Gamma-ray Flaring of BL Lacertae, ApJ, 762, 92

\bibitem[Banasi\'{n}ski et al.(2016)]{banasinski2016}
Banasi\'{n}ski, P., Bednarek, W., \& Sitarek, J. 2016, Orphan $\gamma$-ray Flares from Relativistic Blobs Encountering Luminous Stars, MNRAS, 463, 26

\bibitem[B\"{o}ttcher et al.(2003)]{bottcher2003}
B\"{o}ttcher, M., et al. 2003, Coordinated Multiwavelength Observations of BL Lacertae in 2000, ApJ, 596, 847

\bibitem[B\"{o}ttcher(2005)]{bottcher2005}
B\"{o}ttcher, M. 2005, A Hadronic Synchrotron Mirror Model for the ``Orphan'' TeV Flare in 1ES 1959+650, ApJ, 621, 176

\bibitem[B\"ottcher et al.(2012)]{bottcher2012} B\"ottcher, M., Harris, D. E., \& Krawzcynski, H. (ed.) 2012, Relativistic Jets from Active Galactic Nuclei (Weinheim: Wiley-VCH)

\bibitem[Blandford \& K\"onigl(1979)]{blandford1979} Blandford, R. D. \& K\"onigl, A. 1979, Relativistic Jets as Compact Radio Sources, ApJ, 232, 34
  
\bibitem[Bloom et al.(1997)]{bloom1997}
Bloom, S. D., et al. 1997, Observations of A Correlated Gamma-ray and Optical Flare for BL Lacertae, ApJ, 490, L145

\bibitem[Blumenthal \& Gould(1970)]{blumenthal1970}
Blumenthal, G. R., \& Gould, R. J. 1970, Bremsstrahlung, Synchrotron Radiation, and Compton Scattering of High-Energy Electrons Traversing Dilute Gases, RvMP, 42, 237

\bibitem[Catanese et al.(1997)]{catanese1997}
Catanese, M., et al. 1997, Detection of Gamma Rays with $E>100$ MeV from BL Lacertae, ApJ, 480, 562

\bibitem[Cawthorne et al.(2013)]{cawthorne2013}
Cawthorne, T. V., Jorstad, S. G., \& Marscher, A. P. 2013, Polarization Structure in the Core of 1803+784: A Signature of Recollimation Shocks?, ApJ, 772, 14

\bibitem[Chatterjee et al.(2012)]{chatterjee2012}
Chatterjee, R., et al. 2012, Similarity of the Optical-Infrared and $\gamma$-ray Time Variability of Fermi Blazars, ApJ, 749, 191

\bibitem[Denn et al.(2000)]{denn2000}
Denn, G. R., Mutel, R. L., \& Marscher, A. P. 2000, Very Long Baseline Polarimetry of BL Lacertae, ApJS, 129, 61

\bibitem[Dodson et al.(2017)]{dodson2017}
Dodson, R., Rioja, M. J., Molina, S. N., \& G\'{o}mez, J. L. 2017, High-Precision Astrometric Millimeter Very Long Baseline Interferometry Using A New Method for Multi-Frequency Calibration, ApJ, 834, 177

\bibitem[Formalont(1999)]{formalont1999} Formalont, E. B. 1999, Image Analysis, in: Taylor, G. B., Carilli, C. L., \& Perley, R. A. (eds.), Synthesis Imaging in Radio Astronomy II, ASP Conf. Ser., 180, 301

\bibitem[Fossati et al.(1998)]{fossati1998}
Fossati, G., Maraschi, L., Celotti, A., Comastri, A., \& Ghisellini, G. 1998, A Unifying View of the Spectral Energy Distributions of Blazars, MNRAS, 299, 433
  
\bibitem[Gaur et al.(2015)]{gaur2015}
Gaur, H., et al. 2015, Optical and Radio Variability of BL Lacertae, A\&A, 582, A103

\bibitem[Ghisellini et al.(1998)]{ghisellini1998}
Ghisellini, G., Celotti, A., Fossati, G., Maraschi, L., \& Comastri, A. 1998, A Theoretical Unifying Scheme for Gamma-ray Bright Blazars, MNRAS, 301, 451

\bibitem[Ghisellini et al.(2002)]{ghisellini2002}
Ghisellini, G., Celotti, A., \& Costamante, L. 2002, Low Power BL Lacertae Objects and the Blazar Sequence, A\&A, 386, 833

\bibitem[Giommi et al.(2012)]{giommi2012}
Giommi, P., Padovani, P., Polenta, G., Turriziani, S., D'Elia, V., \& Piranomonte, S. 2012, A Simplified View of Blazars: Clearing the Fog around Long-Standing Selection Effects, MNRAS, 420, 2899

\bibitem[G\'{o}mez et al.(1997)]{gomez1997}
G\'{o}mez, J. L., Mart\'{i}, J. M., Marscher, A. P., Ib\'{a}\~{n}ez, J. M., \& Alberdi, A. 1997, Hydrodynamical Models of Superluminal Sources, ApJ, 482, 33

\bibitem[G\'{o}mez et al.(2016)]{gomez2016}
G\'{o}mez, J. L., et al. 2016, Probing the Innermost Regions of AGN Jets and Their Magnetic Fields with \emph{Radioastron}. I. Imaging BL Lacertae at 21 $\mu$as Resolution, ApJ, 817, 96

\bibitem[Guo et al.(2016)]{guo2016}
Guo, Y. C., Hu, S. M., Li, Y. T., \& Chen, X. 2016, Statistical Analysis of the Temporal Properties of BL Lacertae, MNRAS, 460, 1790

\bibitem[Hodgson(2015)]{hodgson_thesis}
Hodgson, J. A., Ultra-high resolution observations of selected blazars, Ph.D. Thesis, Universit{\"a}t zu K{\"o}ln, 2015

\bibitem[Hodgson et al.(2016)]{hodgson2016}
Hodgson, J. A., Lee, S.-S., Zhao, G.-Y., Algaba, J.-C., Yun, Y., Jung, T., \& Byun, D.-Y. 2016, The Automatic Calibration of Korean VLBI Network Data, JKAS, 49, 137

\bibitem[Hodgson et al.(2017)]{hodgson2017}
Hodgson, J. A., et al. 2017, Location of $\gamma$-ray Emission and Magnetic Field Strengths in OJ 287, A\&A, 597, 80

\bibitem[H\"{o}gbom(1974)]{hogbom1974}
H\"{o}gbom, J. A. 1974, Aperture Synthesis with A Non-Regular Distribution of Interferometer Baselines, A\&AS, 15, 417H

\bibitem[Hovatta et al.(2008)]{hovatta2008}
Hovatta, T., Nieppola, E., Tornikoski, M., Valtaoja, E., Aller, M. F., \& Aller, H. D. 2008, Long-Term Radio Variability of AGN: Flare Characteristics, A\&A, 485, 51

\bibitem[Hovatta et al.(2009)]{hovatta2009}
Hovatta, T., Valtaoja, E., Tornikoski, M., \& L\"ahteenm\"aki, A. 2009, Doppler Factors, Lorentz Factors and Viewing Angles for Quasars, BL Lacertae Objects and Radio Galaxies, A\&A, 494, 527

\bibitem[Hughes et al.(1985)]{hughes1985}
Hughes, P. A., Aller, H. D., Aller, M. F. 1985, Polarized Radio Outbursts in BL Lacertae. The Flux and Polarization of a Piston-Driven Shock, ApJ, 298, 301

\bibitem[Hughes(1991)]{hughes1991} Hughes, P. A. (ed.) 1991, Beams and Jets in Astrophysics (Cambridge: Cambridge University Press)

\bibitem[Jorstad et al.(2001a)]{jorstad2001a}
Jorstad, S. G., Marscher, A. P., Mattox, J. R., Wehrle, A. E., Bloom, S. D., \& Yurchenko, A. V. 2001a, Multiepoch Very Long Baseline Array Observations of EGRET-Detected Quasars and BL Lacertae Objects: Superluminal Motion of Gamma-ray Bright Blazars, ApJS, 134, 181

\bibitem[Jorstad et al.(2001b)]{jorstad2001b}
Jorstad, S. G., Marscher, A. P., Mattox, J. R., Aller, M. F., Aller, H. D., Wehrle, A. E., \& Bloom, S. D. 2001b, Multiepoch Very Long Baseline Array Observations of EGRET-Detected Quasars and BL Lacertae Objects: Connection between Superluminal Ejections and Gamma-ray Flares in Blazars, ApJ, 556, 738

\bibitem[Jorstad et al.(2013)]{jorstad2013}
Jorstad, S. G., et al. 2013, A Tight Connection between Gamma-ray Outbursts and Parsec-Scale Jet Activity in the Quasar 3C 454.3, ApJ, 773, 147

\bibitem[Kadler et al.(2008)]{kadler2008}
Kadler, M., et al. 2008, The Trails of Superluminal Jet Components in 3C 111, ApJ, 680, 867

\bibitem[Kaiser(2006)]{kaiser2006}
Kaiser, C. R. 2006, The Flat Synchrotron Spectra of Partially Self-Absorbed Jets Revisited, MNRAS, 367, 1083

\bibitem[Karamanavis et al.(2016)]{karamanavis2016}
Karamanavis, V., et al. 2016, What Can the 2008/10 Broadband Flare of PKS 1502+106 Tell Us?, A\&A, 590, 48

\bibitem[Kim et al.(2015)]{kim2015}
Kim, J.-Y., et al. 2015, PAGaN I: Multi-Frequency Polarimetry of AGN Jets with KVN, JKAS, 48, 285

\bibitem[Krawczynski et al.(2004)]{krawczynski2004}
Krawczynski, H., et al. 2004, Multiwavelength Observations of Strong Flares from the TeV Blazar 1ES 1959+650, ApJ, 601, 151

\bibitem[Lee et al.(2008)]{lee2008}
Lee, S.-S., et al. 2008, A Global 86 GHz VLBI Survey of Compact Radio Sources, AJ, 136, 159

\bibitem[Lee et al.(2013)]{lee2013}
Lee, S.-S., et al. 2013, Monitoring of Multi-Frequency Polarization of Gamma-ray Bright AGNs, EPJWC, 6107007L

\bibitem[Lee et al.(2014)]{lee2014}
Lee, S.-S., et al. 2014, Early Science with the Korean VLBI Network: Evaluation of System Performance, AJ, 147, 77

\bibitem[Lee et al.(2015)]{lee2015}
Lee, S.-S., et al. 2015, Amplitude Correction Factors of Korean VLBI Network Observations, JKAS, 48, 229

\bibitem[Lee et al.(2016)]{lee2016}
Lee, S.-S., et al. 2016, Interferometric Monitoring of Gamma-ray Bright AGNs. I. The Results of Single-Epoch Multifrequency Observations, ApJS, 227, 8

\bibitem[Le\'{o}n-Tavares et al.(2011)]{tavares2011}
Le\'{o}n-Tavares, J., Valtaoja, E., Tornikoski, M., L\"{a}hteenm\"{a}ki, A., \& Nieppola, E. 2011, The Connection between Gamma-ray Emission and Millimeter Flares in Fermi/LAT Blazars, A\&A, 532, A146

\bibitem[Lewis et al.(2016)]{lewis2016}
Lewis, T. R., Becker, P. A., \& Finke, J. D. 2016, Time-dependent Electron Acceleration in Blazar Transients: X-ray Time Lags and Spectral Formation, ApJ, 824, 108

\bibitem[Lobanov(1998)]{lobanov1998}
Lobanov, A. P. 1998, Ultracompact Jets in Active Galactic Nuclei, A\&A, 330, 79

\bibitem[MacDonald et al.(2015)]{macdonald2015}
MacDonald, N. R., Marscher, A. P., Jorstad, S. G., \& Joshi, M. 2015, Through the Ring of Fire: $\gamma$-ray Variability in Blazars by A Moving Plasmoid Passing A Local Source of Seed Photons, ApJ, 804, 111

\bibitem[Marscher(1995)]{marscher1995}
Marscher, A. P. 1995, Probes of the Inner Jets of Blazars, PNAS, 92, 11439

\bibitem[Marscher et al.(2008)]{marscher2008}
Marscher, A. P., et al. 2008, The Inner Jet of An Active Galactic Nucleus as Revealed by A Radio-to-$\gamma$-ray Outburst, Nature, 452, 966

\bibitem[Marscher et al.(2010)]{marscher2010}
Marscher, A. P., et al. 2010, Probing the Inner Jet of the Quasar PKS 1510-089 with Multi-Waveband Monitoring during Strong Gamma-ray Activity, ApJL, 710, L126

\bibitem[Marscher et al.(2011)]{marscher2011}
Marscher, A. P., Jorstad, S. G., Larionov, V. M., Aller, M. F., \& L\"{a}hteenm\"{a}ki, A. 2011, Multi-Waveband Emission Maps of Blazars, JApA, 32, 233

\bibitem[Marscher(2012)]{marscher2012}
Marscher, A. P. 2012, Structure and Emission of Compact Blazar Jets, IJMPS, 8, 151

\bibitem[Marscher(2014)]{marscher2014}
Marscher, A. P. 2014, Turbulent, Extreme Multi-Zone Model for Simulating Flux and Polarization Variability in Blazars, ApJ, 780, 87

\bibitem[Mart\'{i} et al.(2016)]{marti2016}
Mart\'{i}, J. M., Perucho, M., \& G\'{o}mez, J. L. 2016, The Internal Structure of Overpressured, Magnetized, Relativistic Jets, ApJ, 831, 163

\bibitem[Mart\'{i}-Vidal et al.(2012)]{vidal2012}
Mart\'{i}-Vidal, I., et al. 2012, On the Calibration of Full-Polarization 86 GHz Global VLBI Observations, A\&A, 542, A107

\bibitem[Maselli et al.(2010)]{maselli2010}
Maselli, A., Massaro, E., Nesci, R., Sclavi, S., Rossi, C., \& Giommi, P. 2010, Multifrequency Observations of A Sample of Very Low Frequency Peaked BL Lacertae Objects, A\&A, 512, A74

\bibitem[Max-Moerbeck et al.(2014)]{moerbeck2014}
Max-Moerbeck, W., et al. 2014, Time Correlation between the Radio and Gamma-ray Activity in Blazars and the Production Site of the Gamma-ray Emission, MNRAS, 445, 428

\bibitem[Netzer(2013)]{netzer2013}
Netzer, H. 2013, The Physics and Evolution of Active Galactic Nuclei (New York: Cambridge)

\bibitem[Oh et al.(2015)]{oh2015}
Oh, J., et al. 2015, PAGaN II: The Evolution of AGN Jets on Sub-Parsec Scales, JKAS, 48, 299

\bibitem[Orienti et al.(2007)]{orienti2007}
Orienti, M., Dallacasa, D., \& Stanghellini, C. 2007, Constraining the Nature of High Frequency Peakers, A\&A, 475, 813

\bibitem[Orienti et al.(2013)]{orienti2013}
Orienti, M., et al. 2013, Radio and $\gamma$-ray Follow-Up of the Exceptionally High-Activity State of PKS 1510-089 in 2011, MNRAS, 428, 2418

\bibitem[O'Sullivan \& Gabuzda(2009)]{osullivan2009}
O'Sullivan, S. P., \& Gabuzda, D. C. 2009, Magnetic Field Strength and Spectral Distribution of Six Parsec-Scale Active Galactic Nuclei Jets, MNRAS, 400, 26

\bibitem[Pacholczyk(1970)]{pachol1970}
Pacholczyk, A. G. 1970, Radio Astrophysics (San Francisco: Freeman)

\bibitem[Padovani(2007)]{padovani2007}
Padovani, P. 2007, The Blazar Sequence: Validity and Predictions, Ap\&SS, 309, 63

\bibitem[Park \& Trippe(2017)]{park2016}
Park, J., \& Trippe, S. 2017, The Long-Term Centimeter Variability of Active Galactic Nuclei: A New Relation between Variability Timescale and Accretion Rate, ApJ, 834, 157

\bibitem[Press(1978)]{press1978} Press, W. H. 1978, Flicker Noises in Astronomy and Elsewhere, Comments Astrophys., 7, 103

\bibitem[Raiteri et al.(2013)]{raiteri2013}
Raiteri, C. M., et al. 2013, The Awakening of BL Lacertae: Observations by Fermi, Swift and the GASP-WEBT, MNRAS, 436, 1530

\bibitem[Ramakrishnan et al.(2015)]{ramakrishnan2015}
Ramakrishnan, V., Hovatta, T., Nieppola, E., Tornikoski, M., L\"{a}hteenm\"{a}ki, A., \& Valtaoja, E. 2015, Locating the $\gamma$-ray Emission Site in Fermi/LAT Blazars from Correlation Analysis between 37 GHz Radio and $\gamma$-ray Light Curves, MNRAS, 452, 1280

\bibitem[Ramakrishnan et al.(2016)]{ramakrishnan2016}
Ramakrishnan, V., et al. 2016, Locating the $\gamma$-ray Emission Site in Fermi/LAT blazars - II. Multifrequency Correlations, MNRAS, 456, 171

\bibitem[Rani et al.(2011)]{rani2011}
Rani, B., et al. 2011, Spectral Energy Distribution Variation in BL Lacs and Flat Spectrum Radio Quasars, MNRAS, 417, 1881

\bibitem[Rioja \& Dodson(2011)]{rioja2011}
Rioja, M. J., \& Dodson, R. 2011, High-Precision Astrometric Millimeter Very Long Baseline Interferometry Using A New Method for Atmospheric Calibration, AJ, 141, 114

\bibitem[Rioja et al.(2014)]{rioja2014}
Rioja, M. J., et al. 2014, Verification of the Astrometric Performance of the Korean VLBI Network, Using Comparative SFPR Studies with the VLBA at 14/7 mm, AJ, 148, 84

\bibitem[Rioja et al.(2015)]{rioja2015}
Rioja, M. J., Dodson, R., Jung, T., \& Sohn, B. 2015, The Power of Simultaneous Multifrequency Observations for mm-VLBI: Astrometry Up to 130 GHz with the KVN, AJ, 150, 202

\bibitem[Rybicki \& Lightman(1997)]{rybicki1997}
Rybicki, G. B. \& Lightman, A. P. 1997, Radiative Processes in Astrophysics (Weinheim: Wiley-VCH)

\bibitem[Sandrinelli et al.(2017)]{sandrinelli2017}
Sandrinelli, A., et al. 2017, Gamma-ray and Optical Oscillations of 0716+714, Mrk~421, and BL Lacertae, A\&A, 600, 132

\bibitem[Shepherd et al.(1994)]{shepherd1994}
Shepherd, M. C., Pearson, T. J., \& Taylor, G. B. 1994, Difmap: An Interactive Program for Synthesis Imaging, BAAS, 26, 987

\bibitem[Sokolov et al.(2004)]{sokolov2004}
Sokolov, A., Marscher, A. P., \& McHardy, I. M. 2004, Synchrotron Self-Compton Model for Rapid Nonthermal Flares in Blazars with Frequency-Dependent Time Lags, ApJ, 613, 725

\bibitem[Spada et al.(2001)]{spada2001}
Spada, M., Ghisellini, G., Lazzati, D., \& Celotti, A. 2001, Internal Shocks in the Jets of Radio-Loud Quasars, MNRAS, 325, 1559

\bibitem[Stirling et al.(2003)]{stirling2003}
Stirling, A. M., et al. 2003, Discovery of A Precessing Jet Nozzle in BL Lacertae, MNRAS, 341, 405

\bibitem[Trippe et al.(2011)]{trippe2011}
Trippe, S., et al. 2011, The Long-Term Millimeter Activity of Active Galactic Nuclei, A\&A, 533, A97

\bibitem[Valtaoja et al.(1992)]{valtaoja1992}
Valtaoja, E., \& Ter\"{a}sranta, H., Urpo, S., Nesterov, N. S., Lainela, M., \& Valtonen, M. 1992, Five Years Monitoring of Extragalactic Radio Soures. III. Generalized Shock Models and the Dependence of Variability on Frequencu, A\&A, 254, 71

\bibitem[Valtaoja \& Ter\"{a}sranta(1995)]{valtaoja1995}
Valtaoja, E., \& Ter\"{a}sranta, H. 1995, Gamma Radiation from Radio Shocks in AGN Jets, A\&A, 297, L13

\bibitem[Valtaoja et al.(1999)]{valtaoja1999}
Valtaoja, E., L\"{a}hteenm\"{a}ki, A., Ter\"{a}sranta, H., \& Lainela, M. 1999, Total Flux Density Variations in Extragalactic Radio Sources. I. Decomposition of Variations into Exponential Flares, ApJS, 120, 95

\bibitem[Villata et al.(2004)]{villata2004}
Villata, M., et al. 2004, The WEBT Campaigns on BL Lacertae -- Time and Cross-Correlation analysis of Optical and Radio Lightcurves 1968--2003, A\&A, 424, 497

\bibitem[Wehrle et al.(2016)]{wehrle2016}
Wehrle, A. E., et al. 2016, Erratic Flaring of BL Lac in 2012-2013: Multiwavelength Observations, ApJ, 816, 53

\bibitem[Zhao et al.(2015)]{zhao2015}
Zhao, G.-Y., Jung, T., Dodson, R., Rioja, M., \& Sohn, B. 2015, KVN Source-Frequency Phase-Referencing Observation of 3C 66A and 3C 66B, PKAS, 30, 629


\end{thebibliography}
\end{document}